# Local dynamics of the coffee rust disease and the potential effect of shade

John Vandermeer,[1,2] Pej Rohani, [3] and Ivette Perfecto[2]

1. Department of Ecology and Evolutionary Biology, University of Michigan, Ann Arbor, MI, 48109, USA.
2. School of Natural Resources and Environment, University of Michigan, Ann Arbor, MI, 48109, USA.
3. Odum School of Ecology, University of Georgia, Athens GA 30602, USA.

In this paper we present a mode that incorporates two levels of dynamic structuring of the coffee rust disease (caused by the rust fungus, *Hemileia vastatrix*). First, two distinct spatial scales of transmission dynamics are interrogated with respect to the resultant structure of catastrophic transitions of epidemics. Second the effect of management style, especially the well-known issue of shade management, is added to the base-line model. The final structure includes a simple one-to-one functional structure of disease incidence as a function as well as a classical critical transition structure, and finally a hysteretic loop. These qualitative structures accord well with the recent history of the coffee rust disease.

Coffee is reported to be one of the most traded commodities in the world, (Pendergast, 1999) and the base of economic support for millions of small farmers (Bacon, 2004) and many national economies. The coffee rust disease (Ccaused by the rust fungus, *Hemileia vastatrix*) had been a chronic but relatively mild problem in Latin America, from its appearance in the early 1980s. But, in 2012 an epidemic began extending from Mexico to Perú, with some reports of as much as a 40% - 50% reduction in yield over the region, potentially affecting many coffee producing nations in Latin America (Cressey, 2013). Memories of the devastation caused in south Asia in the nineteenth century (Monaco, 1977; McCook, 2006), causes continuing concern in the region (Schieber and Zentmyer, 1984; Fulton, 1984).

Dealing with the disease is a challenging practical problem, with current recommendations focused on fungicides, search for resistant varieties, and phytosanitation procedures based on uncertain understanding of the ecological dynamics. Indeed, the complexity of this disease has been arguably the reason for failure of most conventional disease control strategies (Avelino et al., 2004; 2012, Kushalappa and Eskes, 1989). Rather, drawing on work by both ecologists (Vandermeer et al., 2009; Jackson et al., 2009; 2012; Soto-Pinto et al., 2002) and phytopathologists (Kushalappa and Eskes, 1989; Avelino et al., 2004; 2012), it could be the larger ecological structure of the agroecosystem that needs to be considered, echoing the many recent calls for a more nuanced approach to the management of ecosystem services in general (Vandermeer and Perfecto, 2012; Perfecto and Vandermeer, 2010; Tscharntke et al., 2012), and pest control specifically (Vandermeer et al., 2010; Lewis et al., 1997). In the present case, specific ecological complications are well-known, forming a complex dynamical system that predicts the possibility of both sustained maintenance of control and occasional escape from control under the same ecological structures (Vandermeer and Rohani, 2014; Vandermeer et al., 2013).

Spatial complexity is acknowledged as an important feature of transmission dynamics in disease ecology (Kitron, 1998; Rohani et al., 1999; Grenfell et al., 2001; Ostfeld et al, 2005; Smith et al., 2005; Riley, 2007). Nevertheless, the consequences of transmission processes acting in potentially different ways at different scales has not been common in the analysis of most disease systems (exceptions include Thrall and Burdon, 1999; Read and Keeling, 2003; 2006; Watts et al., 2005). The medical framework effectively works at the spatial scale of individuals, pursuing the understanding of how a disease spreads through the host body. The classical epidemiological framework focuses at the level of the population, usually elaborating on the mean field approach to the classical susceptible/infected/recovered structure. The landscape level must consider the effect of movement or migration over larger scales, from city to city, from river valley to river valley, and has received very little attention in the disease ecology literature (see for example Kitron, 1998 and Keeling et al. 2010). Transmission processes are clearly distinct within each of these frameworks, yet the majority of disease studies focus on a single spatial scale. Understanding first, the nature of transmission processes specific to each of these scales, and second, the consequences of their interactions, an third, the influence of management strategies may require qualitative theory for educating the intuition of possible general patterns.

Conceptualizing the ecology of transmission as a coupling of distinct spatial scales permits an evaluation of environmental factors as "modifiers" of transmission parameters. Such an approach is especially important in the age of climate change, since many of those modifiers are likely to exert an important effect with expected transformation of climate forcings (Patz, 1996). Our approach in the present work is organized around the ecological factors that modify baseline parameters, specifically effects of management intensity through shade tree density.

There are two spatial scales of the dynamics relevant to the current discussion, transmission among leaves within a plant and transmission from plant to plant on a farm. The relevant equations are (based on Vandermeer and Rohani, 2014):

$$\frac{dx}{dt} = m_1 x(1-x) + m_2 y(1-x) - e_1 x$$

$$\frac{dy}{dt} = m_3 xy(1-y) - e_2 y$$

At equilibrium, we have,

$$y = \frac{e_1 x}{m_2(1-x)} - \frac{m_1 x}{m_2}$$

1a

$$y = 1 - \frac{e_2}{m_3 x}$$

1b

which can be used to analyze the stability conditions as presented in Vandermeer and Rohani (2015).

The question of shade in coffee has long been an important management issue. Specifically with regard to the coffee rust disease, shade does two important things: it reduces wind currents that are responsible for spore dispersal, and it increases humidity, which is necessary for spore germination. Thus, at least theoretically, we might propose that full sun coffee (no shade trees at all) will be inhospitable for the rust because of a lack of moisture (the sun drying out the farm), while full shade is also inhospitable because of lack of wind currents (resulting in limited spore transmission).

Supposing that we can ignore equation 1a (fix x=1) and deal with only equation 1b, the parameter $m_3$ is most likely affected by shade, in a negative (from the rust's point of view) fashion. Thus we might propose:

$m_3 = f_3(S)$

Where S is shade, which here we formulate as ranging from 0 to 1, and we presume,

$$\frac{\partial f_3}{\partial S} < 0$$

whence it is obvious that the equilibrium value of equation 1b becomes,

$$y = 1 - \frac{e_2}{f_3(S)}$$

and,

$$\frac{\partial y}{\partial S} = -\frac{\partial \left[\frac{e_2}{f_3(S)}\right]}{\partial S} = \frac{e_2 \frac{\partial f_3}{\partial S}}{f_3^2}$$

which is always negative, a result that follows elementarily from the assumption that shade reduces wind currents that are responsible for dispersal of spores of the rust. However, if we add equation 1a to the analysis, we assume the reverse relationship with shade for $m_1$, since it is affected mainly by the moisture that comes with shade, such that if $m_1 = f_1(S)$, we assert that,

$$\frac{\partial f_1}{\partial S} > 0.$$

If we take $f_1$ to be an increasing linear function of S, and let $f_3$ be;

$$f_3(S) = a + \frac{bS}{c+S}$$

there appear to be three qualitatively distinct patterns that may form (other than the two trivial solutions where y = 1 for all S or y = 0 for all S). These three forms are illustrated in figure 1, along with the changing structure of the isoclines that produce them. We note the existence of either a hysteretic loop (figure 1c) or a simple critical transition situation (figure 1b), along with what might be the initial expectation of a "humped-shaped" curve associated with the effects of shade.

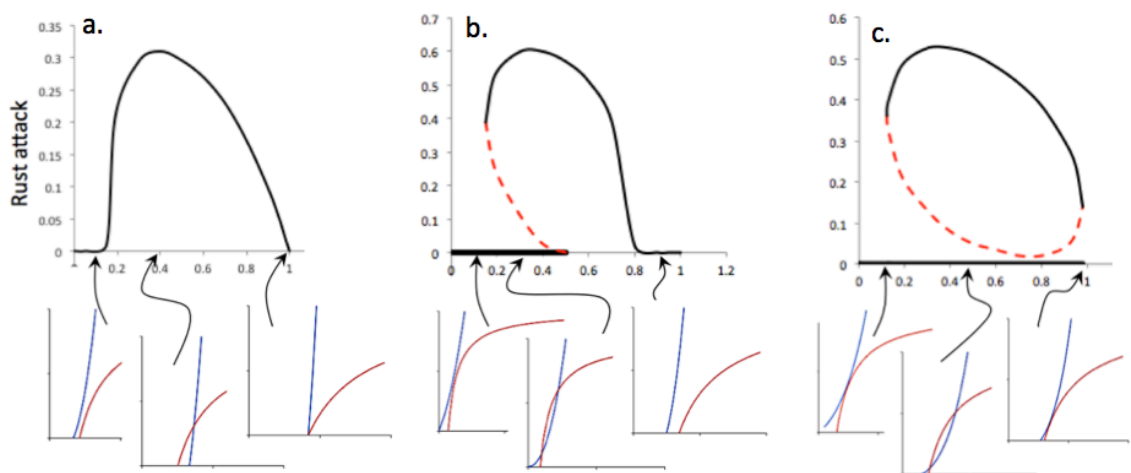

*Figure 1. The three qualitatively distinct outcomes of adding shade to the basic mean field transmission model. Graphs below illustrate the arrangement of the basic model's isoclines at particular points on the x axis, red isocline is equation 1a, blue is equation 1b.*

In sum, this simple model incorporates distinct transmission mechanisms at distinct spatial scales, along with expected patterns associated with shade management in the system. The qualitative patterns that emerge (figure 1) may be thought of as a combination of the "humped-shaped" curve expected from the assumed response of the disease to distinct levels of shade, coupled with the critical transition behavior emerging from two distinct spatial scales of transmission dynamics (Vandermeer and Rohani 2014). It is not unusual for disease systems to contain this sort of behavior (Liu et al., 2012; Earn et al., 2000). The general history of this disease (McCook and Vandermeer, 2015), with a seemingly long-term presence followed by a sudden, and frequently unexpected, epidemic, concords with the qualitative expectations that emerge from this model. The model treats "deforestation" of traditional coffee farms as an effective driver of the critical transition to an epidemic. It is worth noting that landscape level deforestation has also been empirically associated with the incidence of this disease (Avelino et al., 2012), lending support to the idea that shade trees are likely beneficial as buffers against the emergence of epidemics of the coffee rust disease.